\documentclass{ws-procs975x65}
\usepackage{graphicx}
\usepackage{tensor}
\newcommand{\ext}[1]{\mathbf{d}\ensuremath{\mathbf{#1}}}

\newtheorem{thm}{Theorem}[section]
\def\beq{\begin{equation}}
\def\eeq{\end{equation}}

\begin{document}

\title{A study of inhomogeneous massless scalar gauge fields in cosmology}
\author{Ben David Normann$^{1*}$, Sigbj\o rn Hervik$^1$, Angelo Ricciardone$^{1,2}$ and Mikjel Thorsrud$^{3}$}


\address{$^1$Faculty of Science and Technology, University of Stavanger, 4036, Stavanger, Norway\\
$^2$INFN, Sezione di Padova, via Marzolo 8, I-35131, Padova, Italy\\
$^3$Faculty of Engineering, \O stfold University College, 1757 Halden, Norway}
$^*$ Presenter. Correspondance via \textit{ben.d.normann@uis.no}.

\begin{abstract}
Why is the Universe so homogeneous and isotropic? We summarize a general study of a $\gamma$-law perfect fluid alongside an inhomogeneous, massless scalar gauge field (with homogeneous gradient) in anisotropic spaces with General Relativity. The anisotropic matter sector is implemented as a $j$-form (field-strength level), where $j\,\in\,\{1,3\}$, and the spaces studied are Bianchi space-times of solvable type. Wald's no-hair theorem is extended to include the $j$-form case. We highlight three new self-similar space-times: the Edge, the Rope and Wonderland. The latter solution is so far found to exist in the physical state space of types I,II, IV, VI$_0$, VI$_h$, VII$_0$ and VII$_h$, and is a global attractor in I and V. The stability analysis of the other types has not yet been performed. This paper is a summary of ~[\refcite{normann18}],  with some remarks towards new results which will be further laid out in upcoming work.
\end{abstract}


\bodymatter


\section{Introduction}
Why does the Universe seem so isotropic on large scales ~[\refcite{Ade:2013nlj,Ade:2015hxq,Saadeh:2016sak}]? Standard cosmology invokes this observation as a principle, as it is hard to solve the Einstein equations without symmetry requirements~[\refcite{philOfCosm}]. In order to understand this high degree of isotropy within the paradigm of GR, it is necessary to relive the theory of such assumption, seeking to replace it instead by an \textit{explanation}. To this end we have softened the  isotropy requirement of the cosmological principle by studying the Bianchi models of solvable type (types I-VII$_h$). 

A natural generalization of massless scalar fields to the anisotropic case, is the $p$-form field\footnote{We here and throughout reference the form field on the fieldstrength level.} with $p=1$ or $3$. Collectively, we shall refer to this form field as a $j$-form field, since the equations will be the same in either case (as explained in Section \ref{sec:pclass}). The connection to massless scalar fields may be drawn from eq. \eqref{jform}. In this study we summarize some results we have obtained in our study so far, and mention a few results to be further reported on in upcoming work(s)(Normann's Ph.D. thesis). We would like to refer the reader to~[\refcite{normann18}] for further references to previous works on related topics. Also, note the recent works ~[\refcite{almeida18,almeida19}] and many interesting references therein.

\section{The general $p$-form action}
A natural candidate for anisotropic matter sourcing is the one stemming from the general $p$-form action ~[\refcite{thorsrud18}]

\begin{equation}
\label{action}
S_{\rm f}=-\frac{1}{2}\int{\mathcal{P}}\wedge\star\mathbf{\mathcal{P}}\,,
\end{equation}
 where $\mathbf{\mathcal{P}}$ is a $p$ -form constructed by the exterior derivative of a ($p-1$)-form $\mathcal{K}$.
 
 The equations of motion and the Bianchi identity, both obtained from the action, \eref{action}, may now be given as

\begin{align}
&\ext{\mathcal{P}}=0\quad\quad\rightarrow\quad\quad\nabla_{[\alpha_0}\mathcal{P}_{\alpha_1\cdots\alpha_p]}=0\quad\quad\textrm{Bianchi Identity.}\label{dP}\\
&\ext{\star\mathcal{P}}=0\quad\quad\rightarrow\quad\quad\nabla_{\alpha_1}\mathcal{P}^{\alpha_1\cdots\alpha_p}=0\quad\quad\textrm{Equations of motion}\label{StdP}.
\end{align}
Here $\mathcal{P}_{\mu_1\cdots\mu_p}$ are the components of $\mathbf{\mathcal{P}}$ in a general basis. As evident from the above equations\setcounter{footnote}{0}\footnote{the Hodge dual $\mathbf{\star \mathcal{P}}$ is closed}, our study assume that there is no source to the $j$-form field.

\paragraph{General properties of the $p$-form action: } Note that the theories derived from the general $p$-form action \eref{action}
respect the following properties: (i) gauge invariance
$\mathcal{L}\rightarrow\mathcal{L}$ under $\mathbf{\mathcal{K}}\rightarrow\mathbf{\mathcal{K}}+\ext{\mathbf{\mathcal{U}}}$, where $\mathcal{U}$ is a ($p-2$) -form;
(ii) only up to second order derivatives in equations of motion;
(iii) Lagrangian is up to second order in field strength $\mathbf{\mathcal{P}}$;
(iv) constructed by exterior derivative of a $p$-form and 
(v) minimally coupled to gravity.

\section{The $j$-form fluid}
\label{sec:pclass}
In our study, $\mathbf{\mathcal{P}}$ is required homogeneous: $\mathbf{\mathcal{P}}(t,\mathbf{x})\Rightarrow\mathbf{\mathcal{P}}(t)$. However, generally, the underlying gauge field $\mathbf{\mathcal{K}}(t, \mathbf{x})$ is allowed to vary both with space and time. Hence we study a $j$-form fieldstrength with an underlying inhomogenous gauge field. This is different from ~[\cite{Barrow:1996fh}], where the gauge potential is a function of time only. In order to classify the possible cases of $p$-form matter fields that can be constructed from the exterior derivative of a ($p-1$)-form, the following notation is introduced: $\{a,b\}$ where  $a$ denotes the rank of the $p$-form $\mathcal{P}$ and $b$ the rank of its Hodge dual $\star\mathcal{P}$. In four dimensional space-time ($a+b=4)$ there are three distinct cases to consider: (i)$\{2,2\}$, (ii) $\{3,1\}$ or $\{1,3\}$ and (iii) $\{4,0\}$. The degeneracy in (ii) is due to the symmetry of the equations  \eref{dP} and \eref{StdP}\footnote{The reason why this degeneracy is not found in the case (iii) is because $\mathcal{P}\,\neq\,\ext{\mathcal{K}}$ in the case $\{0,4\}$. Hence one is left only with $\{4,0\}$. }. This symmetry can also be seen in the action \eref{action}, up to a prefactor. Since these two cases yield the same set of equations, we effectively study both cases, by studying one of them.

In the following analysis the cases $\{1,3\}$ and $\{3,1\}$ will be taken into account. In order to include both scenarios, notation shall here, and throughout the rest of the paper, be such that $\mathcal{J}$ denotes either (i) the Hodge dual of a 3-form field strength $\mathcal{C}(t)=\ext{\mathcal{B}}$ (where $\mathcal{B}(t,\mathbf{x})$ is a $2$-form) or (ii) the $1$-form field strength $\mathcal{A}(t)=\ext{\phi}$ (where $\phi(t,\mathbf{x})$ is a scalar field). That both cases give rise to the same equations is evident.

\paragraph{Energy-momentum tensor: } The energy-momentum tensor of the $j$-form fluid can be shown to be 
\begin{equation}
\label{jform}
\mathcal{L}_{\rm jf}=-\frac{1}{2}\mathcal{J}_{\mu}\mathcal{J}^{\mu}\quad\rightarrow\quad T_{\mu\nu}^{\rm jf}=\mathcal{J}_{\mu}\mathcal{J}_{\nu}-\frac{1}{2}g_{\mu\nu}\mathcal{J}_\gamma\mathcal{J}^\gamma.
\end{equation}
Equations \eref{dP} and \eref{StdP} now take the component forms
\begin{eqnarray}
\label{dD1}
\ext{\mathcal{J}}&=0\quad\rightarrow\quad\nabla_{[\mu}\mathcal{J}_{\nu]}=0,\\
\label{dStD1}\ext{\star\mathcal{J}}&=0\quad\rightarrow\quad\nabla_{\mu}\mathcal{J}^{\mu}=0.
\end{eqnarray}
These are the equations for a massless scalar field. Hence, our study can be viewed as a study of an inhomogeneous, massless scalar field with a homogeneous gradient.

\paragraph{State parameter} The field strength $\mathcal{J}_{\alpha}$ may be decomposed according to
\begin{equation}
\label{astJ}
\mathcal{J}_\alpha=-w\, u_\alpha+v_\alpha\,,
\end{equation}
where the 4-velocity $u_\alpha$ is time-like ($u_\alpha u^\alpha\,<\,0$), whereas $v_\alpha$ is defined to be orthogonal to $u_\alpha$ and therefore space-like ($v_\alpha v^\alpha\,>\,0$). The range of the equation of state parameter $\xi$ defined through $p_{\rm f}=(\xi-1)\rho_{\rm f}$ \setcounter{footnote}{0}\footnote{in a standard irreducible notation where $p$ denotes pressure and $\rho$ is the energy density. } comes as no surprise. Performing the calculations, one finds
\begin{equation}
\label{eos3f}
\xi=\frac{w^2-v^2/3}{w^2+v^2}+1\phantom{000}\rightarrow\phantom{000} \frac{2}{3}\,\leq\,\xi \,\leq\, 2\,.
\end{equation}
The range of $\xi$ follows directly from requiring that $\mathcal{J}_\alpha\,\in\,\mathbb{R}$. Note that \eref{eos3f} is a dynamical equation of state, since the components of $\mathcal{J}$ in general change with time. The lower bound ($\xi \,=2/3$) is found for $w=0$ and the upper bound ($\xi \,=2$) is found for $v=0$. Note also that $w=v$ gives $\xi \,=\,4/3$, as in the case of electromagnetic radiation.

\section{Sourcing anisotropy with a $j$-form in General Relativity}
\label{sec:formandbianchi}
We take the evolution to be governed by the Einstein Field Equations. In particular
\begin{equation}
\label{Einst}
R_{\mu\nu}-\frac{1}{2}R\,g_{\mu\nu}=T^{\rm pf}_{\mu\nu}+T^{\rm f}_{\mu\nu}+T^{\rm 4f}_{\mu\nu}\,,
\end{equation}
where $R_{\mu\nu}$ is the Ricci tensor components, $R=\tensor{R}{^\mu_\mu}$ is the Ricci scalar and $T^{\rm pf}_{\mu\nu}$ and $T^{\rm f}_{\mu\nu}$ the perfect fluid and 
form fluid energy-momentum tensor components, respectively. The constants $8\pi G$ and $c$ are fixed to 1. Note that a $4$-form is also added, playing the role of a cosmological constant.

It is, for simplicity, assumed that the three fluids do not interact. 
\subsection{Bianchi models and choice of frame}
In dimension three there are nine different (classes of) Lie algebras -- these are the nine different Bianchi types I-IX. The line element of the Bianchi models can be written as
\begin{equation}
{\rm ds}^2=-{\rm d}t^2+\delta_{ab}\,\omega^a\omega^b\phantom{000}\textrm{where}\phantom{000}\ext{\omega}^a=-\frac{1}{2}\tensor{\gamma}{^a_{bc}}\omega^{b}\wedge\omega^c-\gamma^a_{~0c}{\rm d}t\wedge\omega^c.
\end{equation}
 $\{\omega^a\}$ is here a triad of 1-forms, and $\tensor{\gamma}{^a_{bc}}$ are the spatial structure coefficients of the Lie algebra characterizing the corresponding Bianchi type. The tetrad $\{\omega^\alpha\}$ is dual to the vector basis $\{\mathbf{e}_\alpha\}$, which must satisfy the relation 
$[\mathbf{e}_\mu,\mathbf{e}_\nu]=\tensor{\gamma}{^\rho_{\mu\nu}}\mathbf{e}_\rho$. Refer to ~([\refcite{gron07}], Chapter 15) for details. 
The time direction is chosen orthogonal to the orbits of the isometry subgroup (i.e.: orthogonal to the three-dimensional hypersurfaces of homogeneity), and the fundamental observer's 4-velocity is aligned with this direction. It is given by $\mathbf{u}=\frac{\partial}{\partial t}$
where $t$ is the cosmological time. We also define a dimensionless time quantity $\tau$ such that ${\rm d}t/{\rm d}\tau=1/H$, where $H$ is the Hubble parameter. The deceleration parameter $q$ is now such that ${\rm d}H/{\rm d}t=-(1+q)H^2$.

\textbf{A convenient frame} in which to conduct the analysis is the orthonormal frame. Such a frame will give first order evolution equations alongside a set of constraints which are useful to simplify the analysis. 
The Bianchi space-times analyzed in the present paper (I-VII$_h$) admit an Abelian $G_2$ subgroup. This allows for a $1+1+2$ split of the four dimensional space-time. As will become clear later, this translates into a $1+1+2$ decomposition of the Einstein Field Equations, as well as the Jacobi and the Bianchi identities. When the orthonormal frame approach is applied to $G_{2}$ cosmologies, it is common to choose a \textit{group-invariant orbit-aligned} frame, i.e. an orthonormal frame which is invariant under the action of $G_{2}$~[\refcite{dynSys}].

\section{No-hair theorems for the $j$-form}
\label{sec:nohair}
No-hair theorems that in previous literature has been established for the Bianchi space-times in the presence of a cosmological constant and a perfect fluid are in this section extended to the presence of the $j$-form in the equations. In particular, it is found that the cosmic no-hair theorem ~[\refcite{Wald:1983ky}] is valid also in this case \setcounter{footnote}{0}\footnote{Note that an anisotropic fluid may sustain an inflationary phase of expansion if it violates the strong or dominant energy condition ~[\cite{Maleknejad:2012as}]. A $j$-form respects these energy conditions.}. Refer to~[\refcite{normann18}] for proofs.

\begin{thm}[First no-hair theorem]
\label{thmDeSitter}
All Bianchi space-times I-VII$_h$ with a $j$-form, a non-phantom perfect fluid \footnote{A perfect fluid is said to be phantom if $\gamma \,<\,0$. }  and a positive cosmological constant will be asymptotically de Sitter with $\Omega_\Lambda=1$ in the case where $\gamma\,>\,0$ (and similarly $\Omega_\Lambda+\Omega_{\rm pf}=1$ in the case where $\gamma=0$).
\end{thm}

A similar but less general theorem holds also in the case of a perfect fluid with $0\leq\,\gamma\,<\,2/3$:
\begin{thm}[Second no-hair theorem]
\label{NoHair}
All Bianchi space-times I-VII$_h$ with a $j$-form, a non-phantom perfect fluid $\Omega_{\rm pf}$ with equation of state parameter $0\leq\,\gamma\,<\,2/3$ will be asymptotically quasi de Sitter with $q=\frac{3}{2}\gamma-1\,<\,0$.
\end{thm}

\section{New stable, anisotropic, self-similar space-times}
We have also performed a dynamical systems analysis of certain Bianchi types with a $\gamma$-law perfect fluid and a $j$-form fluid. Due to the no-hair theorems, we necessarily remove the cosmological constant in the further study. Among extensions of known self-similar space-times, we interestingly find three new anisotropic space-times; \textit{Wonderland}, \textit{The Edge} and \textit{The Rope} in Bianchi type I. Their global stability has been determined by monotone functions. Extensions to these into other Bianchi types are found, as summarized in the following.

\paragraph{Type I: } Type I splits into two further invariant subspaces; a temporal and a spatial part (as referred to the components of the $j$-form fluid). All of the three new solutions presented below are in the subspace where the $j$-form fluid is purely spatial.

\emph{Wonderland} is an LRS solution containing both a non-rotating vector and the perfect fluid. The field strength is aligned with the LRS axis and the expansion asymmetry is of prolate type. Its range of existence is the open interval $\gamma\in (2/3 ,2)$. It approaches the flat FLRW solution $\Omega_{\rm pf}$ when $\gamma\rightarrow 2/3$ and the Kasner solution ($\Sigma_+=-1$) when $\gamma\rightarrow 2$. Interestingly, it has a deceleration parameter $q=-1+3\gamma/2$ identical to the flat FLRW solution. The line element of Wonderland is
    \begin{equation}
    {\rm d}s^2=-{\rm d}t^2+t^{2}{\rm d}x^2+t^{\frac{2-\gamma}{\gamma}}({\rm d}y^2+{\rm d}z^2).
    \end{equation}
Global attractor for the Bianchi type I state space for $\gamma\,\in\,(2/3,6/5]$.

\emph{The Rope} contains a rotating vector and the perfect fluid.  Its range of existence is the open interval $\gamma\in(6/5, 4/3)$. It approaches Wonderland in the limit $\gamma\rightarrow 6/5$ and the Edge in the limit  $\gamma\rightarrow 4/3$. Like Wonderland, it has a deceleration parameter $q=-1+3\gamma/2$ identical to the flat FLRW solution. The Rope is not an LRS solution, although it is ``almost LRS'' close to Wonderland.  
The line element of the Rope is
    \begin{equation}
    {\rm d}s^2=-{\rm d}t^2+t^2\left({\rm d}x+ \sqrt{\frac{2(5\gamma-6) }{(2-\gamma)}}t^{1-\frac{2}{\gamma}}{\rm d}z\right)^2+t^{\frac{2(4-3\gamma)}{\gamma}}{\rm d}y^2+t^{\frac{4(\gamma-1)}{\gamma}}{\rm d}z^2.
    \end{equation}
Global attractor for the Bianchi type I state space for $\gamma\,\in\,(6/5,4/3)$.
   
\emph{The Edge} contains only a rotating vector and has deceleration parameter $q=1$, similar to a radiation dominated universe. Since $\,\Omega_{{\rm pf}}=0$, it exists in the entire range of models, $\gamma\in[0,2]$. The line element of the Edge is
    \begin{equation}
    {\rm d}s^2=-{\rm d}t^2+t^2\left({\rm d}x+\sqrt{2}t^{-1/2}{\rm d}z\right)^2+{\rm d}y^2+t{\rm d}z^2.
    \end{equation}
 Global attractor for the Bianchi type I state space for  $\gamma\,\in\,[4/3,2)$.
 \paragraph{Type V:} In the type V subspace, we find a one-parameter extension of the  Wonderland equilibrium set. Generally, in this one-parameter family, the $j-$form may have both temporal and spatial components. We also find Plane Waves. For $\gamma\,\in\,(2/3,2)$ these two are the only attractors.
 
\paragraph{Types II, IV, VI$_0$, VI$_h$, VII$_0$ and VII$_h$: } In an upcoming paper (part of Normann's Ph.D. thesis, due in October), we intend to study at least the types I,II,IV,VII$_0$ and VII$_h$ in depth. Here we only report a few preliminary results.

\emph{Wonderland.} This solution (alongside one parameter extensions) are found in the physical state space of the types II,IV VI$_0$, VI$_h$, VII$_0$ and VII$_h$. This is promising in the sense that this solution might not be sensitive to choices of background geometry. There remains, however, some work in order to pin down its stability.

\emph{Edge and Rope.} These two solutions show up also in the physical state space of type II.

\emph{Plane Waves.} In type IV we find a two parameter extension of the plane wave solution. In type VII$_h$, the plane waves is a three parameter family of solutions.

\end{document}